%% file: paper.tex
\documentclass[acmtog,authorversion,nonacm]{acmart}
\usepackage{hyperref}
\usepackage{hyperxmp}
\usepackage[htt]{hyphenat}
\usepackage{graphicx}
\graphicspath{{images/}}
\usepackage[most]{tcolorbox}
\usepackage{xcolor}
\usepackage{soul}
\usepackage{enumerate}
\usepackage{siunitx}
\usepackage{threeparttable}

\definecolor{configuration}{HTML}{000000}
\definecolor{selection}{HTML}{231ECD}
\definecolor{retrieval}{HTML}{F4A658}
\definecolor{augmentation}{HTML}{358C1B}
\definecolor{generation}{HTML}{338791}
\definecolor{background}{HTML}{FFFFFF}

\AtBeginDocument{%
  \providecommand\BibTeX{{%
    \normalfont B\kern-0.5em{\scshape i\kern-0.25em b}\kern-0.8em\TeX}}}

\setcopyright{acmcopyright}
\copyrightyear{2026}
\acmYear{2026}

\acmConference[TScIT 45]{45$^{th}$ Twente Student Conference on IT}{July 3,
  2026}{Enschede, The Netherlands}
\newcommand{\dimension}[3]{%
\begin{tcolorbox}[
    colback=#2!8,
    colframe=#2!80!black,
    boxrule=0pt,
    leftrule=1.5mm,
    sharp corners,
    left=3pt,
    right=3pt,
    top=2pt,
    bottom=2pt,
    boxsep=1pt,
    before skip=2.5mm,
    after skip=0mm
]
\textbf{#1} #3
\end{tcolorbox}
}

\newcommand{\colorunder}[2]{%
  {\setulcolor{#1}\ul{#2}}%
}

\input{sections/commands}

\begin{document}

\title{GuidedRAG: Semantic Steering of Retrieval-Augmented Generation}


\author{Matthijs Jansen op de Haar}
\email{mjanse@student.ethz.ch}
\affiliation{%
  \institution{ETH Zürich}
  \city{Zürich}
  \country{Switzerland}
}

\author{Tobias Stähle}
\email{tobias.staehle@inf.ethz.ch}
\affiliation{%
  \institution{ETH Zürich}
  \city{Zürich}
  \country{Switzerland}
}

\author{Lorenzo Gatti}
\email{l.gatti@utwente.nl}
\affiliation{%
  \institution{University of Twente}
  \city{Enschede}
  \country{The Netherlands}
}

\renewcommand{\shortauthors}{Matthijs Jansen op de Haar, Tobias Stähle and Lorenzo Gatti}

\begin{abstract}

In this work, we propose \textbf{\Framework}, a novel extension to traditional Retrieval-Augmented Generation (RAG) that introduces a dedicated selection stage and semantic steering during retrieval. In contrast to current state-of-the-art RAG approaches, which depend on increasingly complex retrieval and knowledge structures, \Framework\ constrains the knowledge base using semantics before retrieval, aligning the retrieval space with user intent while substantially reducing the search space. Our evaluation shows that \Framework\ improves retrieval relevance by 14.0--15.8\%, mitigates a 19.7--27.4\% loss in retrieval precision, and reduces retrieval overhead by orders of magnitude. Moreover, relevant chunks are consistently retrieved earlier in the ranking process, while alignment with user intent improves by 31.8--36.8\%. We further show that \Framework\ achieves full coverage across 15 diverse RAG variants, demonstrating generalizability across the literature. Together, these findings establish semantic steering and selections as a powerful and generalizable paradigm for improving the current state-of-the-art in RAG.

\end{abstract}




\settopmatter{printacmref=false}

\begin{teaserfigure}
  \includegraphics[width=\textwidth]{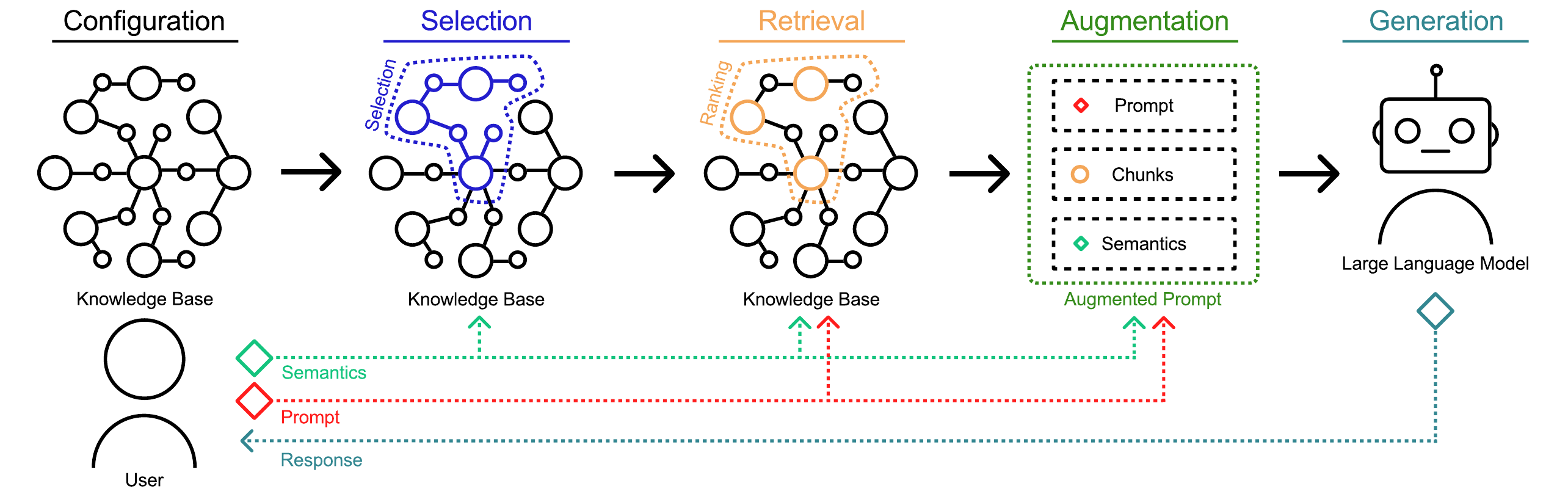}
  \caption{In \Framework\ we investigate five stages, namely: The \colorunder{configuration}{Configuration}, \colorunder{selection}{Selection}, \colorunder{retrieval}{Retrieval}, \colorunder{augmentation}{Augmentation}, and \colorunder{generation}{Generation} stage. The \colorunder{configuration}{Configuration} stage specifies the knowledge base (KB) and underlying semantics, and configures the degree of user control. In the Retrieval-Augmented Generation (RAG) process, a user specifies \textit{semantics} and a \textit{prompt}, which are used in subsequent steps. In the \colorunder{selection}{Selection} stage, a subset of the KB is created based on the provided semantics. This subset is then used to apply ranking in the \colorunder{retrieval}{Retrieval} stage to find relevant chunks. The user-specified prompt, semantics and retrieved chunks are then augmented into a new prompt in the \colorunder{augmentation}{Augmentation} stage, after which a Large Language Model (LLM) generates a response in the \colorunder{generation}{Generation} stage.}
  \Description{Pipeline describing GuidedRAG}
  \label{fig:teaser}
\end{teaserfigure}

\maketitle

\input{sections/01_introduction}

\input{sections/02_background}

\input{sections/03_framework}

\input{sections/04_methodology}

\input{sections/05_evaluation}

\input{sections/06_discussion}

\input{sections/07_conclusion}


\bibliographystyle{ACM-Reference-Format}
\bibliography{references}

\clearpage

\appendix

\section{Comparison of RAG approaches}
\begin{table}[h]
\begin{threeparttable}

\centering
\caption{This table presents a mapping of 15 distinct RAG approaches that fit inside the \Framework\ framework. Each approach is mapped across the \Framework\ stages, with each stage corresponding to a method in the given RAG approach. The mapping covers each approach exhaustively, resulting in full coverage of the corpus. Therefore, \Framework\ can extend any of the listed RAG approaches, and generally extends to any other RAG approach that makes use of the canonical \colorunder{retrieval}{Retrieval},  \colorunder{augmentation}{Augmentation}, and \colorunder{generation}{Generation} stages in their implementation. In addition, \Framework\ supports any knowledge base in the \colorunder{configuration}{Configuration} stage, as long as it contains semantics and chunks. The latter being a prerequisite for any RAG framework \cite{rag, graphrag, RAG_survey, ir_llms_survey}.}
\label{tab:mapping}
\begin{tabular}{ccccccc}
\toprule
Index  & RAG Approach $\ddagger$ & \colorunder{configuration}{Configuration} & \colorunder{selection}{Selection} &\colorunder{retrieval}{Retrieval} & \colorunder{augmentation}{Augmentation} $\dagger$ & \colorunder{generation}{Generation} \\
\midrule
1& RAG \cite{rag}& VectorDB & N/A & Dense Retrieval & N/A & LLM\\
2& GraphRAG \cite{graphrag}& KG & Entity Tagger & Hybrid Retrieval & Relations & LLM\\
3& LightRAG \cite{lightrag} & KG + VectorDB & Entity Tagger & Hybrid Retrieval & Relations + Entities & LLM\\
4& FiD \cite{FiD} & VectorDB & N/A & Dense Retrieval & N/A & Fusion-in-Decoder \\
5& Self-RAG \cite{selfrag} & VectorDB & N/A & Dense Retrieval & Reflection Tokens & LLM\\
6& REALM \cite{realm} & VectorDB & N/A & Dense Retrieval & N/A & BERT\\
7& RAFT \cite{raft} & VectorDB & N/A & Dense Retrieval & N/A & Fine-Tuned LLM\\
8& HippoRAG \cite{hippo} & KG & Entity Tagger & Graph Retrieval & Relations & LLM\\
9& FLARE \cite{flare} & VectorDB & N/A & Active Retrieval & Predicted Tokens & LLM \\
10& DPR \cite{DPR} & VectorDB & N/A & Dense Retrieval & N/A & N/A $\S$ \\
11& RETRO \cite{retro} & VectorDB & N/A & Nearest-Neighbor & N/A & Transformer Decoder\\
12& Atlas \cite{rag_bloat} & VectorDB & N/A & Dense Retrieval & N/A & LLM \\
13& FiD-KD \cite{FiD} & VectorDB & N/A & Dense Retrieval & N/A & Fusion-in-Decoder \\
14& Query2Doc \cite{query2doc} & VectorDB & N/A & Dense Retrieval & N/A & LLM \\
15& Adaptive-RAG \cite{adaptive_RAG} & VectorDB & N/A & Adaptive Retrieval & Complexity Signal & LLM\\
\bottomrule
\end{tabular}

\begin{tablenotes}
\item[$\dagger$]{\ Chunks are omitted, as these are always present in the augmentation stage}
\item[$\ddagger$]{\ We include approaches that do not rely on LLMs for generation}
\item[$\S$]{\ No explicit generation model}
\end{tablenotes}

\end{threeparttable}

\end{table}

\clearpage

\section{GuidedRAG WORKFLOW EXAMPLE}

\noindent
\begin{minipage}{\textwidth}
    \centering
    \includegraphics[width=\textwidth]{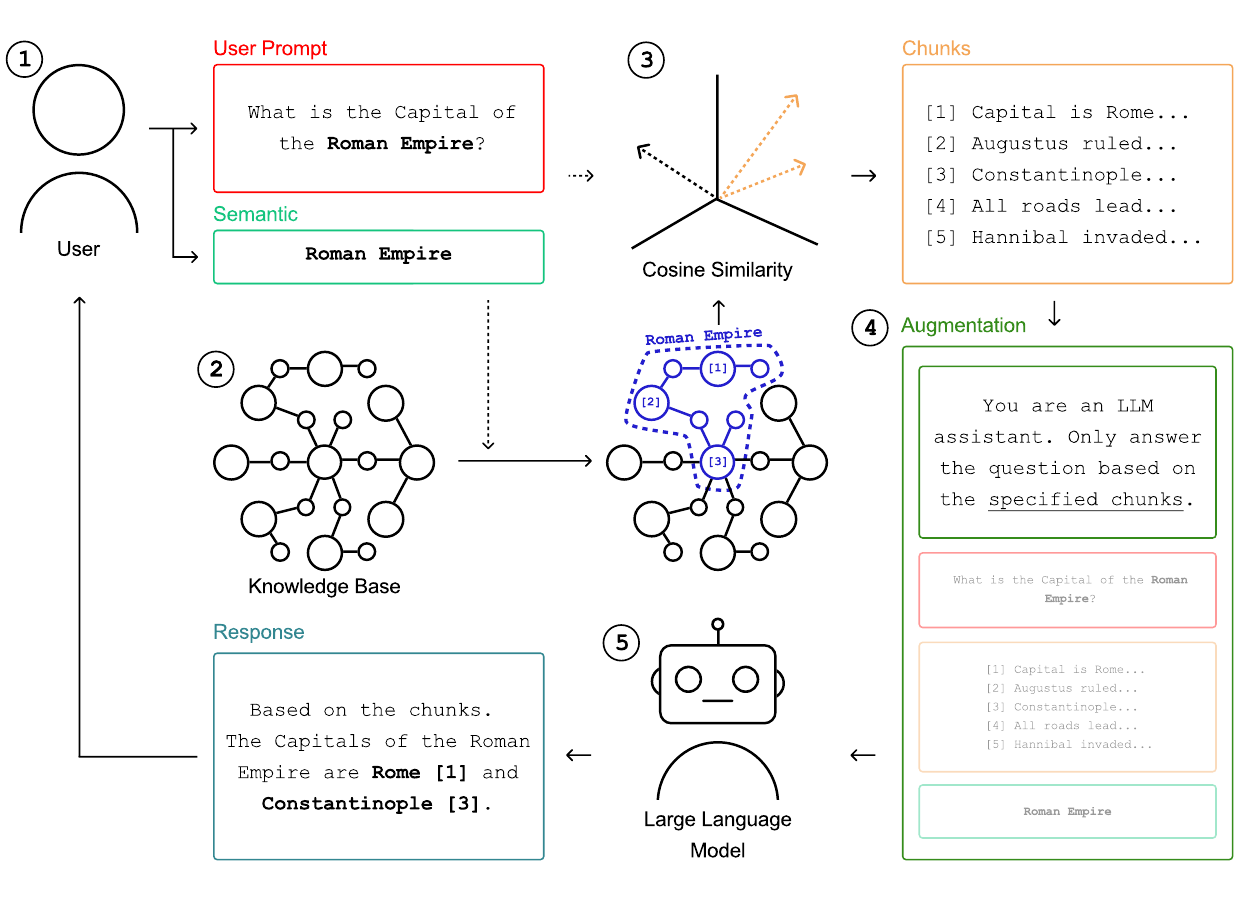}
    \captionof{figure}{In this example, we illustrate an instance of using \Framework. \textbf{(i)} During the \colorunder{configuration}{Configuration} stage, a user specifies a prompt and a semantic. In this case, the question concerns the \texttt{Roman Empire}. \textbf{(ii)} Using this semantic, the knowledge base is scoped in the \colorunder{selection}{Selection} stage to include only chunks related to the \texttt{Roman Empire}. \textbf{(iii)} This selection space is then used to rank the chunks according to their similarity to the user prompt, using cosine similarity as the \colorunder{retrieval}{Retrieval} method. \textbf{(iv)} The user prompt, retrieved chunks, and semantics are then combined with a system prompt in the \colorunder{augmentation}{Augmentation} stage, resulting in an augmented prompt. \textbf{(v)} Finally, the augmented prompt is used in the \colorunder{generation}{Generation} stage and sent to an LLM, which generates a response that is sent to the user.} 
    \label{fig:altr_teaser}
\end{minipage}

\clearpage

\section{GuidedRAG Extended Examples}
In this section, we provide some additional examples of \Framework, namely: the \colorunder{configuration}{Configuration}, \colorunder{selection}{Selection}, \colorunder{retrieval}{Retrieval}, \colorunder{augmentation}{Augmentation}, and \colorunder{generation}{Generation} stages contextualized in concrete examples of systems in which \Framework\ can be used; to best illustrate where semantic steering and selections can come into play. 

\def\Map{\texttt{GooMaps}}
\def\Legal{\texttt{AceWright}}

\subsection{Map Example}
In this example, we discuss \Map, a \textit{fictional} navigation and map-based system that contains various locations, restaurants, and stores (i.e., entities). The application can be used to explore these entities and to summarize reviews and blog posts associated with them.

\dimension{Configuration:}{configuration}{In \Map, information is stored about individual locations, restaurants, and stores, including reviews and blog posts. This information is represented using a vector database, where chunks are associated with specific entities (e.g., a restaurant) and additional categories (e.g., cuisine). A user interacts with \Map\ by selecting an entity on the map, such as a restaurant. Upon selection, the system can generate a summary of the restaurant using \Framework. Additionally, the user may ask specific questions, such as details about menu items.}

\dimension{Selection:}{selection}{After a restaurant has been selected, the selection stage defines a selection space over the relevant chunks. In \Map, this consists of chunks associated only with the selected restaurant, such as its reviews, descriptions, and related blog content. By restricting the selection space to a single entity, the system reduces the search scope while maintaining relevance, without requiring additional knowledge-base filtering.}

\dimension{Retrieval:}{retrieval}{During retrieval, the scoped selection space is used to support both summarization and question answering. For summarization, maximal marginal relevance is applied to retrieve a diverse set of chunks that capture different aspects of the restaurant (e.g., food quality, service, and atmosphere). For question answering, cosine similarity is used to retrieve the most relevant chunks with respect to the user query, such as a question about vegetarian options or pricing.}

\dimension{Augmentation:}{augmentation}{After retrieving the relevant chunks, the augmentation layer combines them with the user query to construct a system prompt. In \Map, this prompt explicitly includes the restaurant name and a base description, ensuring that the generation stage remains grounded in the selected entity. This additional context helps the model distinguish between similar entities and improves response specificity.}

\dimension{Generation:}{generation}{Finally, the response is generated using an LLM. For summarization, the model produces a coherent overview of the restaurant by integrating information from multiple chunks. For question answering, the model generates a focused response based on the retrieved evidence. In both cases, restricting the output to entity-specific chunks ensures the generated output remains accurate and relevant to the selected restaurant.}

\subsection{Legal Example}
In this example, we discuss \Legal, a \textit{fictional} legal information system that contains various laws, regulations, and legal cases (i.e., entities). The application can be used to explore these entities, summarize legal documents, and answer legal questions.

\dimension{Configuration:}{configuration}{The configuration stage describes the knowledge base and user-system interactions. \textbf{(i)} In \Legal, information is stored about legal entities such as statutes, case law, and regulations, including their descriptions, interpretations, and references. This information is represented using a combination of a knowledge graph and a vector database, where chunks are associated with specific entities (e.g., a law or case) and relations (e.g., \textit{cites}, \textit{amends}, or \textit{applies to}). \textbf{(ii)} A user interacts with \Legal\ by selecting a legal entity or submitting a legal query, such as a specific law or legal topic. Upon selection, the system can generate a summary of the selected law using \Framework. Additionally, the user may ask targeted questions, such as how a law applies in a specific scenario or which cases reference it.}

\dimension{Selection:}{selection}{After a legal entity has been selected, the selection stage defines a selection space over the relevant chunks. In \Legal, this consists of chunks associated with the selected law or case, as well as closely related entities connected through the knowledge graph (e.g., referenced cases or amended statutes). By restricting the selection space to the selected entity and its immediate relations, the system reduces the search scope while preserving legally relevant context.}

\dimension{Retrieval:}{retrieval}{During retrieval, the scoped selection space is used to support both summarization and question answering. For summarization, maximal marginal relevance is applied to retrieve a diverse set of chunks capturing different aspects of the legal entity (e.g., definitions, applications, and interpretations). For question answering, cosine similarity is used to retrieve the most relevant chunks with respect to the user query, such as a question about the applicability of a law in a specific case.}

\dimension{Augmentation:}{augmentation}{After retrieving the relevant chunks, the augmentation layer combines them with the user query to construct a system prompt. In \Legal, this prompt explicitly includes the name of the legal entity (e.g., a statute) and a base description, ensuring that the generation stage remains grounded in the correct legal context. Additionally, references to related entities may be included to provide supporting context for interpretation.}

\dimension{Generation:}{generation}{Finally, the response is generated using an LLM. For summarization, the model produces a coherent overview of the legal entity by integrating information from multiple chunks, such as its purpose and scope. For question answering, the model generates a focused response based on the retrieved evidence, such as explaining how a law applies to a given scenario. In both cases, restricting the output to entity- and relation-specific chunks ensures that the generated response remains accurate, grounded, and legally consistent.}

\section{Prompt}
This section details the prompt used to mark a given chunk as relevant during the evaluation. Annotations were averaged over three separate iterations, due to the inherent non-determinism of large language models, even when the temperature is set to zero. 

\dimension{Relevance Annotation:}{generation}{\\
You are the best LLM-based annotator. You are evaluating whether a chunk retrieved through RAG contains a fact; if it does, then it is deemed relevant. Return ONLY valid JSON that contains a boolean indicating whether a chunk was relevant (true) or not (false).\\

Make sure that the chunk explicitly contains the fact, as otherwise you should return false.\\ 

FACT:\\
\{ fact \}\\

QUESTION:\\
\{ question \}\\

CHUNK:\\
\{ chunk \}.\\

OUTPUT FORMAT:\\
\{ "relevant": boolean \}}

\vspace{30mm}
\end{document}

%% file: sections/commands.tex
\def\Framework{\textcolor{black}{\texttt{GuidedRAG}}}
\def\FrameworkRAG{\textcolor{black}{\textbf{RAG}}}
\def\FrameworkGraphRAG{\textcolor{black}{\textbf{GraphRAG}}}

\def\Dataset{\textcolor{black}{\texttt{ARLtR}}}

%% file: sections/01_introduction.tex
\input{sections/commands}
\section{Introduction}
The rapid advancement of \textit{Large Language Models (LLMs)} has contributed significantly to natural language understanding and generation across a wide range of applications, including \textit{Information Retrieval (IR) Systems} \cite{opportunities_llms_in_ir, on_llms_in_ir, ir_llms_survey}. However, despite their impressive capabilities, LLMs remain limited by their static training data and susceptibility to generating factually incorrect or outdated information \cite{rag, few_shot_learners}. \textit{Retrieval-Augmented Generation (RAG)} seeks to mitigate these issues by augmenting prompts with external knowledge retrieved from a knowledge base at inference time, thereby improving factual accuracy and adherence to domain-specific knowledge \cite{rag}.

While RAG has demonstrated substantial benefits, its effectiveness is fundamentally dependent on the quality and relevance of the retrieval process \cite{RAG_survey, rag}. Traditional RAG \cite{rag} systems often rely on broad semantic similarity searches over large document collections, which can introduce irrelevant context, increase computational costs, and reduce retrieval precision \cite{distraction_effect, zhao2024retrievalaugmentedgenerationaigeneratedcontent}. As knowledge bases continue to grow in size and complexity, these limitations become increasingly pronounced, negatively impacting both efficiency and response quality. Furthermore, approaches such as \textit{GraphRAG} \cite{graphrag} rely heavily on generated entity annotations. As a result, inaccuracies or omissions in these annotations can introduce a semantic gap between the user's intent and the knowledge retrieved \cite{tagging_inaccuracies}.

To address these challenges, we propose \Framework\ which extends traditional RAG approaches by introducing a \textbf{(i)} selection stage and \textbf{(ii)} semantic steering during the canonical RAG process. In the selection stage, a knowledge base (KB) is scoped to form a selection space using semantic information, such as categories or entities. This reduces retrieval overhead by narrowing the search scope, thereby making RAG systems more scalable. Furthermore, it improves retrieval relevance by grounding the search space in high-level semantics. As a result, retrieval focuses on semantically relevant portions of the corpus, reducing the inclusion of irrelevant context. Moreover, while standard approaches rely solely on a user prompt to steer retrieval, \Framework\ improves user control by allowing pre-specified semantics to steer retrieval. This results in closer alignment between user intent and machine interpretation. In this work, we demonstrate how \Framework\ can improve efficiency, accuracy, and precision versus traditional RAG approaches. 

\textbf{Main RQ:} How does \Framework\ compare to state-of-the-art retrieval-augmented generation approaches in terms of \textbf{(i)} retrieval relevance, \textbf{(ii)} retrieval efficiency, and \textbf{(iii)} retrieval precision?

\begin{enumerate}[(i)]
    \item \textbf{Retrieval Relevance:} How does the relevance and accuracy of retrieved context compare against \textit{RAG} and \textit{GraphRAG}?

    \item \textbf{Retrieval Efficiency:} How much do semantic selections improve retrieval efficiency and ranking of relevant context?

    \item \textbf{Retrieval Precision:} To what extent is the inclusion of irrelevant context reduced compared to \textit{RAG} and \textit{GraphRAG}?
\end{enumerate}

%% file: sections/02_background.tex
\input{sections/commands}
\section{Background}

LLMs are neural networks, specifically transformers \cite{attention_is_all_you_need}, trained on large collections of text to perform natural language understanding and generation tasks. During training, knowledge is encoded in the model's parameters, enabling it to generate responses based on patterns learned from the training corpus. While this approach has proven highly effective, it also introduces several limitations. Since knowledge is stored in pre-trained parameters, LLMs cannot directly access information unavailable during training \cite{rag, roberts2020knowledgepackparameterslanguage}. Furthermore, domain-specific knowledge may be underrepresented, resulting in inaccurate or incomplete responses \cite{rag}. This can additionally result in hallucinations, where LLMs make up facts or claims \cite{hallucinations}. These limitations motivate the use of external knowledge sources during inference, thereby grounding knowledge and further augmenting the context window and understanding of LLMs \cite{rag}. 

Retrieval-Augmented Generation (RAG) \cite{rag} addresses the limitations of static knowledge by retrieving information from an external KB and incorporating it alongside a user-generated query provided to an LLM. As a result, generated responses can be grounded in information beyond the model's pre-trained parameters \cite{rag}. Traditional RAG systems rely on embedding models and dense retrieval. Documents, i.e., chunks of text, are transformed into embeddings and stored in a (vector) database, while user queries are converted into embeddings using the same embedding model according to the vector space model \cite{vector_space_model}. Relevant chunks are then retrieved by comparing embedding similarity \cite{rag}. In recent years, the definition of what constitutes a RAG system has grown to include alternative approaches such as \textit{GraphRAG} \cite{graphrag}. These approaches generally change one or multiple components in the canonical RAG process of retrieval, augmentation, and generation. In GraphRAG, a knowledge graph (KG) \cite{hogan2021knowledge} is used to provide relevant semantic context during retrieval to incorporate relational context during the generation stage. In this work, we will evaluate \Framework\ against traditional RAG and GraphRAG, as these are two popular yet distinct approaches that are commonly cited in the literature and used in industry \cite{RAG_survey, han2025rag}. Although RAG certainly improves access to external knowledge, retrieval quality often declines as the knowledge base grows \cite{rag_bloat}. Larger search spaces increase computational costs and can introduce irrelevant and misleading context into the retrieval process \cite{rag_bloat}. Furthermore, retrieval noise can cascade into the generation stage, leading to hallucinations \cite{hallucinations}, incorrect or flawed responses, and increased token usage, all of which persist even with large context windows \cite{du-etal-2025-context,rag}.

The effectiveness of a RAG-based system is largely determined by the quality of its retrieval process \cite{rag,rag_bloat}. Since generated responses depend on the retrieved context as input, retrieval should provide information that is both relevant and efficient to obtain \cite{rag_bloat, rag}. Retrieving too much information can cause information to be diluted or the context window to overflow, while retrieving too little might cause relevant context to be lost during inference \cite{du-etal-2025-context,attention_is_all_you_need}. Therefore, it is important to evaluate strategies that achieve higher retrieval relevance and that score relevant chunks more favorably. This will allow for the inclusion of fewer chunks during augmentation, reducing token usage and generation delays. Furthermore, retrieval should remain resilient in the face of ever-growing knowledge bases while retrieving context that closely aligns with user intent \cite{rag_bloat}. Improving accuracy alone is not sufficient; precision is also important to avoid hallucinations when information is absent from the KB \cite{rag,rag_bloat}. These requirements motivate the use of a selection stage in \Framework; this stage narrows the KB to include semantically relevant regions during retrieval. This reduces retrieval overhead, as embedding similarity is compared over a smaller corpus, and improves response quality by considering only semantically relevant chunks during retrieval. As semantics are grounded in the KB, it would be possible to determine whether relevant chunks exist, improving precision. Furthermore, by leveraging semantics, the canonical RAG process can be steered to employ retrieval or augmentation techniques tailored to the scenario, such as vector similarity \cite{vector_space_model} or maximal marginal relevance \cite{mmr}. The use of semantic steering and selections remains underexplored in existing RAG literature and the current state-of-the-art approaches.

%% file: sections/03_framework.tex
\input{sections/commands}
\section{Framework}
We present \Framework\ as illustrated in Figure \ref{fig:teaser}. Adding a \colorunder{configuration}{Configuration} and \colorunder{selection}{Selection} stage that facilitates semantic steering in the canonical \colorunder{retrieval}{Retrieval}, \colorunder{augmentation}{Augmentation}, and \colorunder{generation}{Generation} stages.

\subsection{Configuration}
The \colorunder{configuration}{Configuration} stage defines the KB and user interactions within a RAG system. KBs can take different forms, such as vector databases that use the vector space model \cite{vector_space_model}, or knowledge graphs \cite{hogan2021knowledge} that contain entities, chunks, and their relations. Moreover, this stage defines the prompt and semantics that are used in subsequent stages, resulting in a selection over the KB or adjustments in retrieval and augmentation approaches. In general, the prompt and semantics are generated by a user. However, depending on system requirements, semantics can be automatically deduced from a prompt or alternatively inferred from system interactions. Therefore, prompt creation and semantic inference are reliant on a balance between user expertise and control. The semantics can serve multiple roles within the framework, including defining retrieval scopes through KB metadata, guiding the retrieval stage toward relevant information, and enriching the augmented prompt with supplementary context.

\dimension{Example:}{configuration}{A user specifies a prompt containing a factoid question and two semantics: \textit{Entity: Rome}, \textit{Question Type: Factoid}.}

\subsection{Selection}
\begin{figure}[h]
    \centering
    \includegraphics[width=1\linewidth]{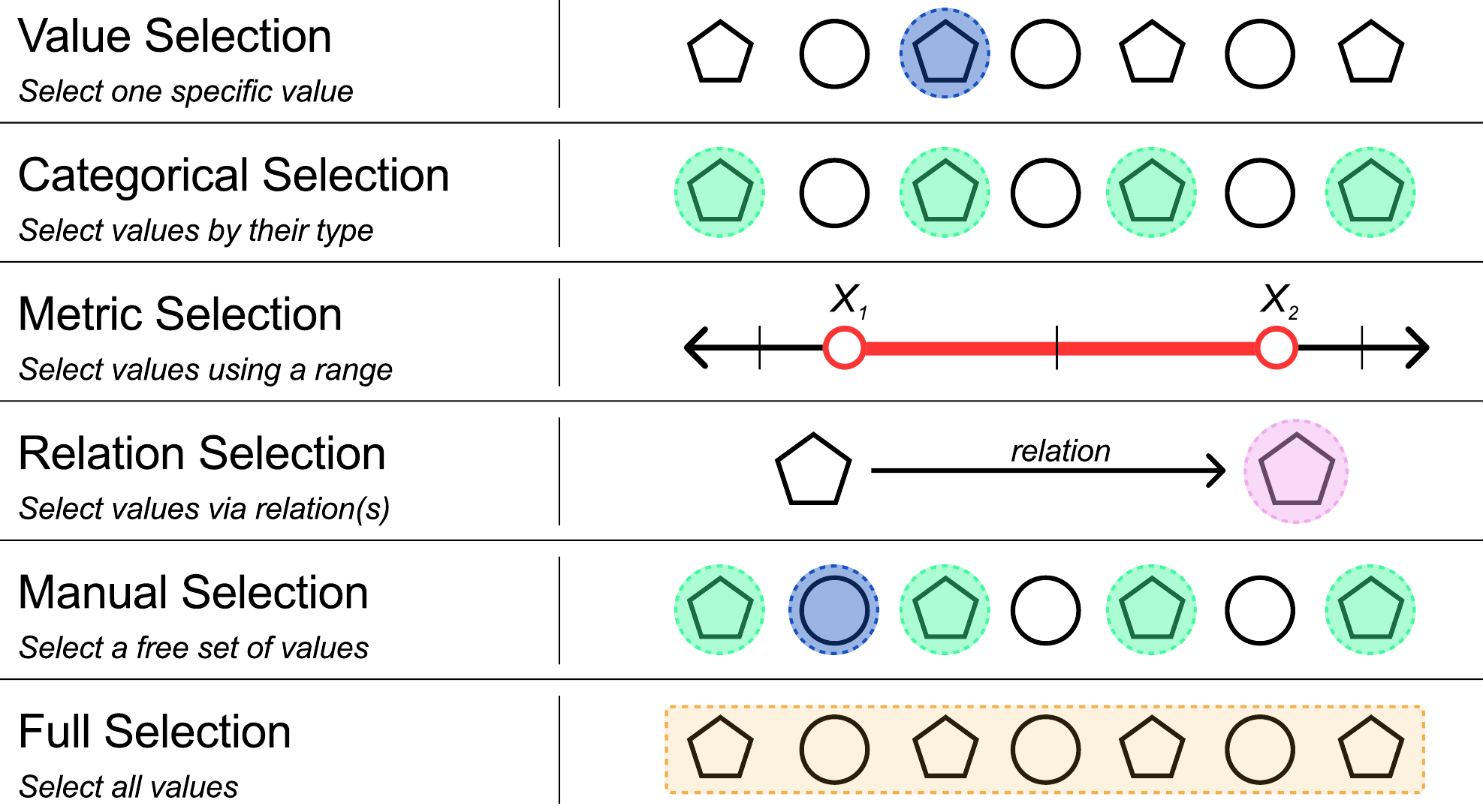}
    \vspace{-4mm}
    \caption{Selection methods use different semantics (e.g., entities or relations) to create subsets over a KB, allowing for retrieval from semantically relevant regions of the KB. For example, selecting chunks from \texttt{2002--2026 AD}.}
    \label{fig:selection methods}
    \vspace{-3mm}
\end{figure}

In the \colorunder{selection}{Selection} stage, semantics are used to create a selection over the entire KB. This selection then creates a subset of the KB used in subsequent stages. Depending on the selection, the corpus is narrowed, improving efficiency and relevance during retrieval. A user or system may also specify no selection, in which case the entire KB is used. As a consequence, the system uses the underlying RAG approach as is. Semantics used for selection can be explicitly specified by a user or determined through user interactions within a system. In the latter case, a system may guide a user to create suitable selections. A selection may consist of one or multiple selection methods; these methods have been defined in Figure \ref{fig:selection methods}, grounded in IR literature \cite{query_retrieval, relational_selection}, and can be used to create a selection space. Similar to query languages, such as SQL, different selections form sets that can be manipulated using \textit{mathematical set theory} operations, such as: union \((\cup)\), intersection \((\cap)\), symmetric difference \((\triangle)\), etc. 

\dimension{Example:}{selection}{The selection stage finds all relevant chunks related to \textit{Entity: Rome}, drastically reducing the search space.}

\subsection{Retrieval}
For the \colorunder{retrieval}{Retrieval} stage, various retrieval methods can be used, depending on the underlying RAG approach. If a selection was specified, a given retrieval technique will rank relevant chunks within the selection space, rather than across the entire KB. This makes retrieval far more efficient, depending on the specificity of the selection, as ranking metrics are computed over a smaller corpus. In traditional RAG methods, ranking techniques are statically configured and rely exclusively on a user prompt as input. In contrast, in \Framework, semantics can be used to switch ranking techniques. For example, an open question may benefit from a diverse ranking method such as Maximum Marginal Relevance (MMR) \cite{mmr}, whereas a factoid question may be better served by a similarity-based ranking method \cite{vector_space_model}. Additionally, the semantics can make ranking more diverse by enabling the use of new ranking techniques. This results in potential improvements in retrieval performance. A powerful consequence of incorporating semantics into the retrieval process is the ability to determine whether information is actually present within the KB. When the specified semantics cannot be matched with equivalent semantics in the KB, the retrieval stage can infer that the relevant information is likely absent. This improves precision and reduces the number of false positives that would have otherwise been retrieved. In traditional RAG approaches, retrieval mechanisms typically return the most similar chunks based on similarity scores, even when no truly relevant information exists \cite{selective_answering, rag_bloat}.

\dimension{Example:}{retrieval}{Based on the specified \textit{Question Type: Factoid}, the retrieval stage uses \textit{cosine similarity} to retrieve relevant chunks.}

\subsection{Augmentation}
In the \colorunder{augmentation}{Augmentation} stage, chunks from the retrieval stage are augmented, along with a user query, into a system-specified prompt format. The chunks are used in the later generation stage to ground responses in the provided context, improving faithfulness \cite{tamber-etal-2025-benchmarking}. In \Framework, semantics can additionally be added to the augmented prompt to provide further context during the generation stage. This approach is already used in RAG frameworks such as GraphRAG \cite{graphrag}, where relational triplets are augmented into a prompt. In contrast, \Framework\ allows for any specification of semantics, generalizing this approach. Furthermore, the semantics can also be used to employ different prompting strategies. This results in different prompt formats, or LLMs, enabling better task descriptions and the use of more suitable models for specific use-cases or requirements.

\dimension{Example:}{augmentation}{The augmentation stage combines the prompt, chunks, and semantics, providing relevant additional context.}

\subsection{Generation}
Finally, in the \colorunder{generation}{Generation} stage, a response is generated using the augmented prompt and returned to a user. In traditional RAG methods, a response may indicate what chunks were utilized to improve transparency. Using the semantics in \Framework, it would be further possible to provide additional context, such as sources or metadata from the retrieved chunks. Furthermore, as the reduction in the search space removes irrelevant context, chunks are retrieved earlier in rankings. This results in potentially fewer chunks in the augmented prompt, reducing response time and token usage.

\dimension{Example:}{generation}{The generation stage responds to the user, using the augmented prompt. As the selection and retrieval resulted in fewer, but more relevant chunks, the response was generated more quickly and accurately, and consumed fewer tokens.}

%% file: sections/04_methodology.tex
\input{sections/commands}
\section{Methodology}

\subsection{Dataset}
In order to validate \Framework\ against base RAG \cite{rag} and GraphRAG \cite{graphrag}, it is essential to use a knowledge base that supports both graph- and vector-based retrieval. Furthermore, the knowledge base must contain textual chunks, entities, and relations, as well as an extensive set of questions paired with corresponding ground truths (i.e., answers). For the evaluation of \Framework, these questions must also be accompanied by semantic annotations to enable appropriate use of the selection stage and semantic steering. The current state-of-the-art dataset, given the constraints, is \textit{All Relations Lead to Rome} (\Dataset), created by Jansen op de Haar et al. \cite{all_roads_lead_to_rome}. \Dataset\ is a knowledge graph stored in Neo4j\footnote{https://neo4j.com/} containing chunks, entities, and relations that have been annotated by an LLM. The dataset contains data from 300 Wikipedia articles about the \textbf{Roman Empire}. 

\Dataset\ includes 6,000 answerable question-answer pairs, denoted with \(Q_a\). Each question has a set of attributes that can be used in the evaluation, as illustrated in Table \ref{tab:question_configurations}. These attributes serve as semantics to define a selection space during retrieval. Questions are categorized as either open-ended or factoids. In the latter case, a unique ground-truth answer exists, enabling relevance assessment of retrieved chunks. Furthermore, the answers are grounded in the KB; so each question is linked to a specific entity, such as \textit{Rome} or \textit{Hannibal}. The \textit{complexity} attribute defines the number of entities (i.e., one (\(C_1\)) or two (\(C_2\))) in a question, while the \textit{relational} attribute indicates whether it concerns an entity or its relation. The latter is particularly important when evaluating GraphRAG \cite{graphrag}, as it generally performs better on multi-hop and relation-centric questions \cite{han2025rag, graphrag}. Moreover, each \textit{base question} has three formulations that correspond to levels of user expertise. This distinction is commonly used in the field of \textit{Human-Computer Interaction (HCI)} \cite{user_expertise} and provides a better representation of user expertise levels. \Dataset\ includes an additional 2,400 unanswerable question-answer pairs about semantics that are not present in the KB, denoted with \(Q_{n}\). This facilitates precision evaluation, as RAG approaches may incorrectly mark chunks as relevant, leading to false positives.  

\begin{table}[h]
  \caption{\Dataset\ Question Attributes}
  \label{tab:question_configurations}
  \centering
  \vspace{-1.5mm}
  \begin{tabular}{ccc}
    \toprule
    Attribute & \(\#\) & Configuration Types \\
    \midrule
     Question Type & 2 & Factoid \ | \ Open \\
     Relational & 2 & Entity (\(E\)) \ | \ Entity + Relation \\
     Complexity & 2 & Single (\(C_1\)) \ | \ Double (\(C_2\)) \\
     User Expertise & 3 & Novice \ | \ Intermediate \ | \ Expert \\
    \bottomrule
  \end{tabular}
  \vspace{-3mm}
\end{table}

\subsection{RAG Methods}
\begin{figure}[t]
    \centering
    \includegraphics[width=1\linewidth]{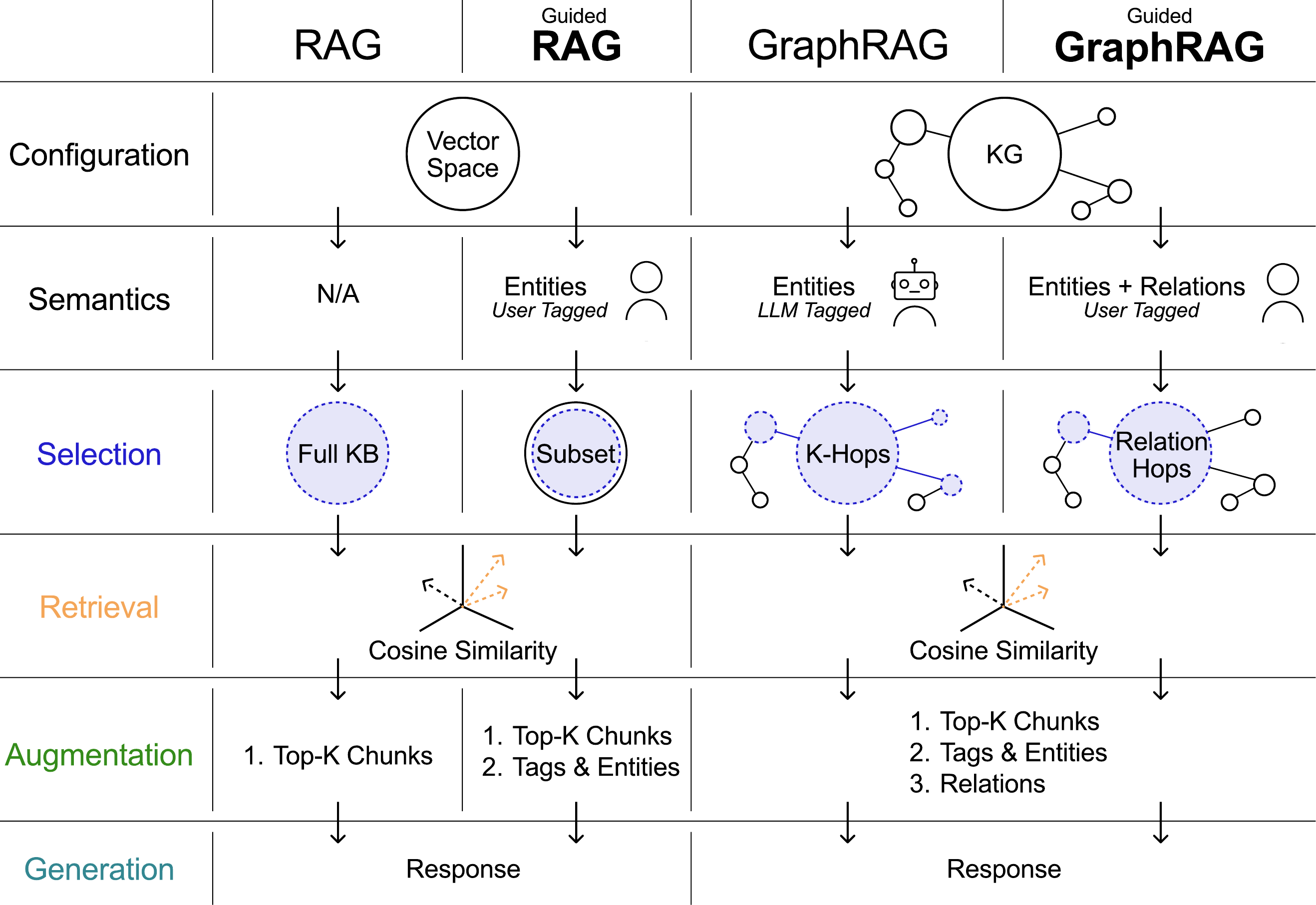}
    \vspace{-5mm}
    \caption{RAG Methods over \Framework\ Framework Stages}
    \label{fig:llm_implementation_table}
    \vspace{-5mm}
\end{figure}

To compare traditional RAG approaches against \Framework, we implement base RAG \cite{rag} and GraphRAG \cite{graphrag}, two widely used approaches in the literature, along with their corresponding \Framework\ counterparts. The implementation for each approach, corresponding to the \Framework\ stages, is described in Figure \ref{fig:llm_implementation_table}. To improve reproducibility, each approach uses cosine similarity \cite{vector_space_model} for retrieval to reduce variance caused by the choice of retrieval method rather than the influence of semantic steering. Traditional RAG \cite{rag} makes use of a vector-based KB and cosine similarity to find the \(top-k\) relevant chunks, where \(k\) defines the number of chunks retrieved. Its \Framework\ counterpart uses a pre-specified entity (\(E\)) to narrow the corpus before applying cosine similarity. Meanwhile, for GraphRAG \cite{graphrag}, an entity is tagged using an LLM or Embedding Model (EM), after which the tagged entity is matched to one in the KB. This entity is then used to apply \(k-hops\); expanding the selection space via connected entities to include relevant additional context, where \(k\) is the number of hops taken. In contrast, the \Framework\ counterpart to GraphRAG uses predefined entities and relations, allowing relation hops that expand the selection space over specific relation types rather than all relations. These relation hops are often infeasible in traditional GraphRAG systems because inaccuracies in entity taggers can introduce irrelevant context into the selection space. In the implementation and evaluation, we use an LLM (i.e., \textit{minimax 2.7}\footnote{https://www.minimax.io/models/text/m27}) and an EM (i.e., \textit{gemini-embedding-2}\footnote{https://deepmind.google/models/gemini/embedding/}).

\subsection{Selection and Retrieval}
The evaluation will use a best-effort strategy, depending on the RAG approach, to showcase performance across the various question configurations in \Dataset\, as set out in Table \ref{tab:question_configurations}. Furthermore, as performance in the augmentation and generation stages largely depends on the choice of LLM, we will primarily evaluate the selection and retrieval stages. These stages will best illustrate the impact of introducing \Framework, in particular on retrieval performance. 

In addition, \Dataset\ provides entities and relations that define the question configuration. Using this information, we use the \textit{complexity} and \textit{relational} attributes to determine which operations are used to construct a selection space. Each configuration, and its subsequent selection space, is shown in Table \ref{tab:retrieval methods}. The selection spaces are defined using \textit{mathematical set theory}. In base RAG, the entire corpus is selected, which is denoted by \(U\). The \Framework\ counterpart makes use of pre-specified entities to scope the corpus, denoted by \(E_n\), where \(n\) denotes the \(n_{th}\) entity. In the event of a Double (\(C_2\)) Entity configuration, the \Framework\ counterpart uses the intersection operation \((\cap)\), as said chunks will contain contextual information about both entities. Furthermore, when using GraphRAG, a similar selection is used for the Entity questions. However, we use a more conservative strategy, employing the union operation \((\cup)\), for the \(C_2\) Entity configuration. This is because incorrect entity tagging may obstruct relevant context. For relational questions, the \(k-hop\) strategy is used, in which the selection space is expanded from a tagged entity to include neighbors linked by relations; these neighbors are denoted by the set \(A\). In the \Framework\ counterpart, rather than using \(k-hops\), relational hops can be used to expand the selection space according to specific relation types rather than all relations. 

\begin{table}[h]
  \caption{Retrieval Configurations and Corresponding Selections}
  \label{tab:retrieval methods}
  \centering
  \vspace{-2mm}
  \begin{tabular}{ccccc}
    \toprule
    Configuration & RAG & \FrameworkRAG* & GraphRAG & \FrameworkGraphRAG* \\
    \midrule
    \(C_1\) Entity & \(U\) & \(E_1\) & \(E_{1}\) & \(E_1\) \\
    \(C_2\) Entity & \(U\) & \(E_1\cap E_2\) & \(E_{1}\cup E_{2}\) & \(E_1\cap E_2\) \\
    \(C_1\) Relation & \(U\) & \(E_1\) & \(E_{1} \cup A_{hop}\) & \(E_1 \cap A_{rel}\) \\
    \(C_2\) Relation & \(U\) & \(U\) & \(E_{1} \cup A_{2*hop}\) & \(A_{rel}\) \\
    \bottomrule
    &&&\multicolumn{2}{c}{* \Framework\ variants.} 
  \end{tabular}
  \vspace{-6mm}
\end{table}

%% file: sections/05_evaluation.tex
\input{sections/commands}
\section{Evaluation}

\subsection{Entity Tagging}
In order to compare the different RAG approaches, we first examine the impact of using automated entity taggers versus pre-specified semantics. The former are widely used in approaches such as GraphRAG \cite{graphrag}. In our comparison, we investigate three approaches: an \textbf{(i)} \textit{LLM-based tagger}, \textbf{(ii)} \textit{cosine similarity}, and a \textbf{(iii)} \textit{hybrid approach} in which LLM-generated tags are matched using an EM (i.e, with 3,072 dimensions). The evaluation will include 300 open questions and 300 factoid questions, with an even split over the \textit{complexity} attribute. This results in a total of 900 entities to be tagged, since \(C_2\) questions are paired with two entities. 

The results are shown in Table \ref{tab:tagging_methods}. We observe that a naive \textit{cosine similarity} tagging strategy severely underperforms compared to the other two methods. Meanwhile, the \textit{LLM-based tagging} strategies perform better, with slightly over half of the tags correctly identified. The combined method achieves the highest performance because the standard LLM tagger may use incorrect entity formulations, which are corrected during the cosine similarity step. Nevertheless, when compared against the ground truth, entities are often misaligned. This is a direct result of question formulations, as entities might not always be explicitly mentioned or, in rare cases, might be formulated incorrectly, such as with the novice user type. It is important to note that users may also choose suboptimal semantics, leading to suboptimal selections. Nevertheless, in automated tagging methods, a discrepancy between user intent and a written prompt can arise, a problem that \Framework\ avoids because semantics are pre-specified. 


\begin{figure}[h]
    \vspace{-2mm}
    \centering
    \includegraphics[width=1\linewidth]{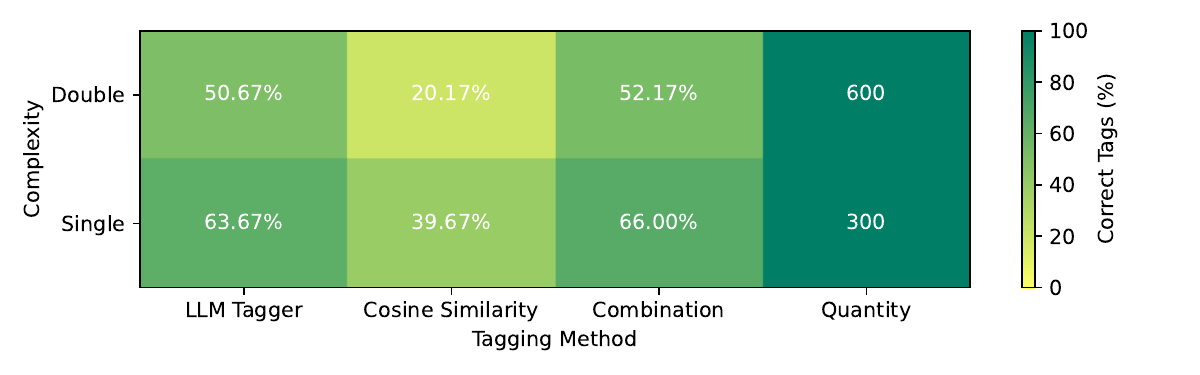}
    \vspace{-7mm}
    \caption{Percentage of Correct Tags per Tagging Method}
    \label{fig:tagging_comparison}
    \vspace{-4mm}
\end{figure}

In addition, automated tagging strategies are computationally expensive because they require LLMs and EMs. This increases computational complexity, as shown in Table \ref{tab:tagging_methods}, resulting in greater token usage and overhead that would otherwise be avoided with pre-specified semantics. Furthermore, in Figure \ref{fig:tagging_comparison}, we observe that as the complexity attribute (\(C\)) increases, the automated tagging strategies perform sub-optimally. This is particularly true for \textit{cosine similarity}, which otherwise has a respectable performance for \(C_1\) type questions. The poor performance is due to using \textit{cosine similarity} on a single question to determine multiple entities; often shifting focus on a single entity rather than identifying multiple entities. 

\begin{table}[h]
  \vspace{-1.5mm}
  \caption{Complexity and Accuracy over Tagging Methods}
  \vspace{-2mm}
  \label{tab:tagging_methods}
  \centering
  \begin{tabular}{ccccc}
    \toprule
    Tagging Type & LLM & EM & Retrieval & Acc. \\
    \midrule
    Cosine Similarity Tagger & - & \(O(1)\) &\(O(n)\) & 26.7\%\\
    LLM-Based Tagger & \(O(1)\) & - &\(O(1)\) & 55.0\%\\
    Combination & \(O(1)\) & \(O(1)\) &\(O(n)\) & 56.8\%\\
    Pre-Specified Tags & - & - &\(O(1)\) & 100.0\%\\
    \bottomrule
  \end{tabular}
  \vspace{-1mm}
\end{table}

\subsection{Retrieval Relevance}
To evaluate retrieval relevance, we examine the performance of the four RAG approaches, as set out in Figure \ref{fig:llm_implementation_table}, on a subset of 600 factoid questions from \(Q_a\). Each factoid has a corresponding \textit{fact} (i.e., answer), grounded in the KB. Utilizing this \textit{fact}, we can determine whether a chunk is relevant to answer a question in the generation stage. In comparing the RAG approaches, we evaluate whether each approach can correctly retrieve the \textit{fact} from at least one of five chunks per question (i.e., \(top-k\) chunks, where \(k=5\)). This results in 12,000 total chunks; each having its relevance annotated as a Boolean value by an LLM at a temperature of zero. These annotations are kept consistent across each RAG approach for a given \textit{base question}. 

Furthermore, entity semantics are taken directly from \Dataset\ rather than being automatically tagged prior to retrieval. This distinction mitigates the influence of entity-tagging variations, which would otherwise confound comparison. The annotations were averaged over three separate iterations to reduce variance in the tagging process, as the inference process is non-deterministic. In addition, to validate the annotation procedure, a representative subset of the LLM annotations, comprising 20\% of the corpus, was independently annotated by the author, yielding a \textit{Cohen's} \(\kappa\) of 92.8\% \cite{cohen_kappa}. 

\begin{table}[h]
  \vspace{-1mm}
  \caption{Top-K Hit Statistics}
  \vspace{-2mm}
  \label{tab:top_k_table}
  \centering
  \begin{tabular}{cccc}
    \toprule
    Approach &Avg. Hit (k) & Median Hit (k) & Relevant (\%) \\
    \midrule
    RAG &3.52&3.0& 59.7\%\\
    \FrameworkRAG* &\textbf{3.08}&\textbf{2.0}& \textbf{68.2\%}\\
    GraphRAG &3.32 &2.5& 66.1\%\\
    \FrameworkGraphRAG* &\textbf{2.71}&\textbf{1.0}& \textbf{76.6\%}\\
    \bottomrule
    &&\multicolumn{2}{c}{* \Framework\ variants.} 
  \end{tabular}
  \vspace{-2mm}
\end{table}

\begin{figure}[h]
    \vspace{-2mm}
    \centering
    \includegraphics[width=1\linewidth]{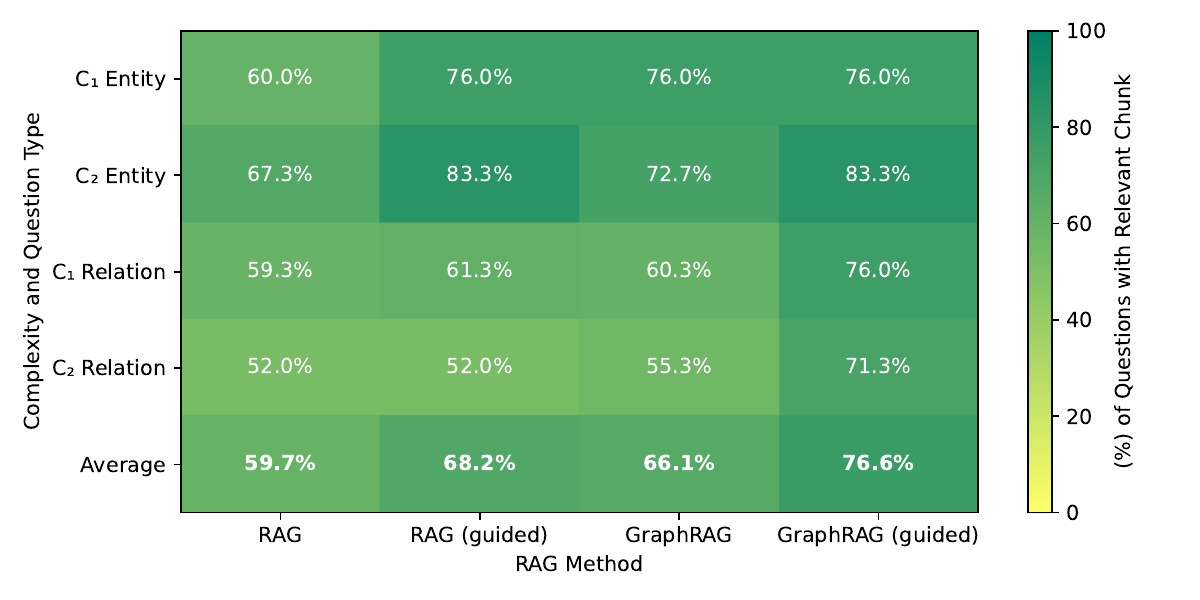}
    \vspace{-7mm}
    \caption{Percentage of Questions with at Least One Correct Chunk}
    \label{fig:chunk_relevance}
    \vspace{-3mm}
\end{figure}

In Figure \ref{fig:chunk_relevance}, the percentage of answerable questions (i.e., using a \textit{fact}) per configuration is shown. We observe that \Framework\ approaches perform considerably better than their traditional counterparts and even outperform state-of-the-art approaches such as GraphRAG \cite{graphrag}. Interestingly, the \Framework\ variant of base RAG performs better than GraphRAG, despite the latter being highly resource-intensive and requiring a knowledge graph. The addition of a selection space yields substantial improvements for \textit{entity} type questions, particularly those of high complexity, where multiple entities may mislead the \textit{cosine similarity} ranking. Interestingly, only the \Framework\ variant of GraphRAG does well on the \textit{relational} type questions, while regular GraphRAG underperforms on this metric. This is due to the \(k-hop\) strategy, which expands the search scope over all relation types, barely narrowing the search space. However, it should be noted that much of the performance in GraphRAG is a result of the augmentation stage, where relations are added into a prompt \cite{graphrag, han2025rag}. Therefore, even if the retrieval stage does not retrieve the \textit{fact}, the relations augmented into the prompt may still yield a correct response during generation. On the contrary, it is important to note that the evaluation uses ground-truth entity annotations, which favor GraphRAG. Overall, \Framework\ performs much better and retrieves more relevant context. In particular, the \Framework\ variant of GraphRAG performs extremely well on the dataset by fully leveraging the provided semantics.

Table \ref{tab:top_k_table} illustrates the average and median \(k_{th}\) position in the \(top-k\) ranking where a given \textit{fact} was retrieved (i.e., \textit{hit}). Each metric accounts for questions where no \textit{fact} was identified, which are weighted as having a \textit{fact} at \(k=6\), favoring larger search spaces and thereby traditional RAG approaches. We observe that the \Framework\ approaches encounter a \textit{fact} much higher in the \(top-k\) ranking when compared to their traditional counterparts. 

In \Dataset\ all \textit{facts} are grounded in the KB, which means that relevant chunks will eventually appear as the \(k\) parameter in \(top-k\) ranking increases to include the entire KB (\(U\)). Hence, systems often improve recall by increasing \(k\) and retrieving more chunks, which provide additional context. However, this comes at the cost of token usage and computational complexity, as the generation stage will require an LLM to process a larger input prompt. Moreover, despite large context windows, this can lead to inaccurate responses \cite{du-etal-2025-context}, or the \textit{lost in the middle} effect, in which LLMs neglect the middle section of a given prompt \cite{lost_in_middle}. \Framework\ mitigates these issues, as relevant chunks are, on average, retrieved higher in the similarity ranking. This allows systems to handle a lower \(k\) parameter in the \(top-k\) retrieval. This results in reduced computational overhead, improved accuracy, and decreased token usage during generation.

\subsection{Retrieval Efficiency}
To assess retrieval efficiency, we investigate the reduction in search space \((S)\) during retrieval resulting from the introduction of a selection stage. The minimum, average, and maximum search space, for each selection in Table \ref{tab:retrieval methods}, are depicted in Table \ref{tab:search_space_per_operation}. The total number of chunks in \Dataset\ is 16,069, which represents the search space when using base RAG; as \textit{cosine similarity} is applied across the entire corpus \((U)\). The results indicate a severe reduction in the search space across different selections, decreasing the computational overhead by orders of magnitude. The use of \Framework\ yields very specific subsets of the KB, improving recall and making retrieval substantially faster, as embedding distances are calculated over a fraction of the corpus. In addition, unlike traditional RAG approaches, embedding computations are dependent on the average number of chunks per selection space (e.g., per entity), rather than the corpus size, thereby making \Framework\ scalable even with much larger and complex corpora. Furthermore, when using \(k-hops\), we observe that with the increase of \(k\), the scope of the search space rapidly increases. This is an emergent property in graphs caused by the low degree of separation between any two entities, also known as the \textit{small world problem} \cite{small_world_problem}. Consequently, this reduces the impact of automated tagging misalignment. However, it results in a sizable search space and additional overhead for \textit{GraphRAG} during retrieval, which likely explains its moderate retrieval relevance. It is important to note that \Dataset\ is a dense knowledge graph; therefore, the average search space is higher than in sparser graphs. 

\begin{table}[h]
  \vspace{-2mm}
  \caption{Search Spaces per Selection}
  \label{tab:search_space_per_operation}
  \vspace{-2mm}
  \centering
  \begin{tabular}{lrrrr}
    \toprule
    Selection & Min. \(S\) & Avg. \(S\) &  Max. \(S\) & Avg. \(S\) \% of \(U\)  \\
    \midrule
    \(U\) & 16,069 & 16,069.00  & 16,069& 100.00\%\\
    \(E_1\) & 1 & 4.57  & 3,476& 0.03\%\\
    \(E_1 \cup E_2\) & 1 & 9.10  & 4,169& 0.06\%\\
    \(E_1 \cap E_2\) & 1 & 1.50  &  432&0.01\%\\
    \(E_1 \cup A_{hop}\)  & 1 & 1,915.53  & 13,856& 11.92\%\\
    \(E_1 \cup A_{2*hop}\) & 1 & 12,149.27  & 14,178& 75.61\%\\
    \(E_1 \cup A_{rel}\) & 1 & 40.80  & 10,787& 0.25\%\\
    \(E_1 \cap A_{rel}\) & 1 & 0.63 &2,825&0.00\%\\
    \(A_{rel}\) & 1 & 36.69 & 10,136& 0.23\% \\
    \bottomrule
  \end{tabular}
  \vspace{-1mm}
\end{table}

\begin{figure}[h]
    \centering
    \vspace{-2.5mm}
    \includegraphics[width=1\linewidth]{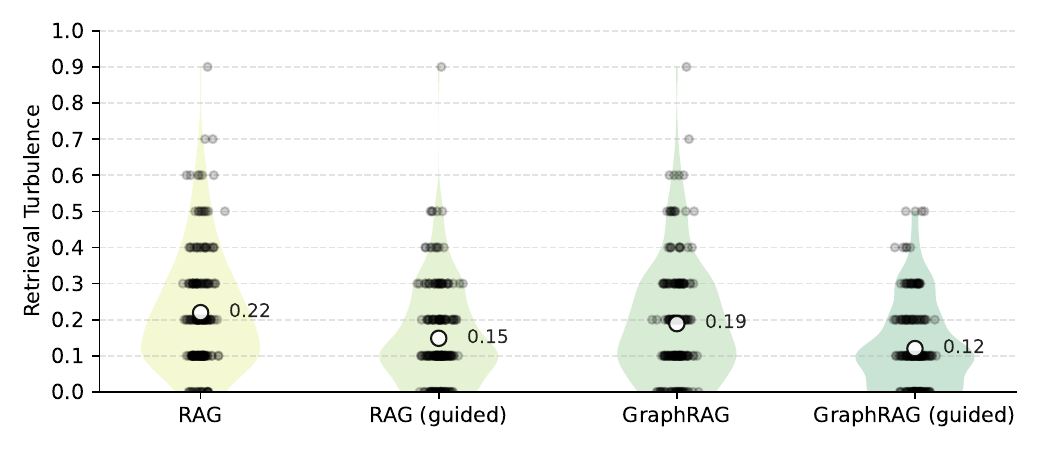}
    \vspace{-7mm}
    \caption{Turbulence of Chunks in Retrieval (0 = low, 1 = high)}
    \label{fig:turbulence}
    \vspace{-2.5mm}
\end{figure}

In Figure \ref{fig:turbulence}, we report retrieval turbulence $(T)$ per RAG approach across three paraphrased formulations \((P)\) of 200 distinct \textit{base questions} from \(Q_a\), which denote question intent. Each formulation corresponds to a given level of user experience, as shown in Table \ref{tab:question_configurations}. Each formulation is provided in \Dataset\ and has \(top-k\) retrieved chunks, where \(k=5\). Turbulence is computed as a function of the number of distinct retrieved chunks across these formulations, denoted $D_c$.
\begin{equation}
T = \frac{D_c - k}{(P\cdot k)-k} \quad\quad or \quad\quad T = \frac{D_c - 5}{10}
\end{equation}

Interestingly, the \Framework\ variants have substantially reduced turbulence across different formulation levels compared to their traditional counterparts. This suggests that differences in user expertise have a smaller effect on retrieval, and further supports our claim that \Framework\ more closely aligns with user intent. This is most certainly a consequence of the reduced search space, as small differences in similarity scoring are less likely to change the ranking results. Nevertheless, this shows great promise in facilitating usage across different levels of user expertise and intent alignment, as retrieval is less dependent on prompt formulation and user expertise.

\subsection{Retrieval Precision}
\begin{figure}[h]
    \vspace{-2mm}
    \centering
    \includegraphics[width=1\linewidth]{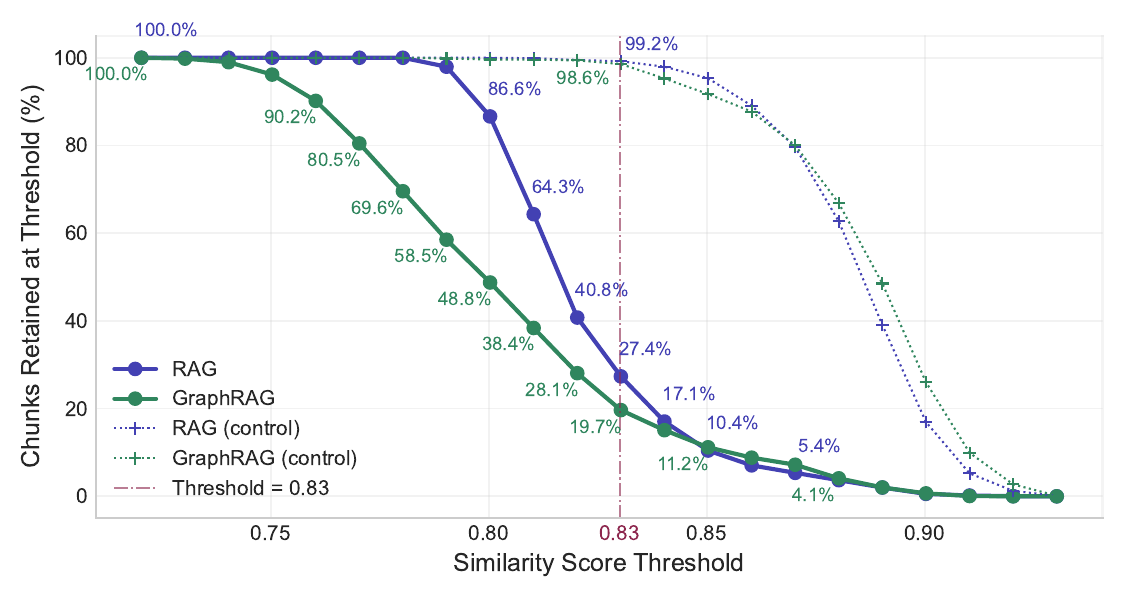}
    \vspace{-7mm}
    \caption{Cumulative Chunks Retained at Similarity Threshold (\%)}
    \label{fig:relevance}
    \vspace{-2mm}
\end{figure}

Retrieval precision is evaluated using a representative sample of 540 unanswerable questions on non-existent semantics sampled from \(Q_n\). In traditional RAG systems, chunk relevance is determined using a cosine similarity score that decides the \(top-k\) ranking. Additionally, a threshold score is used to determine whether a question is answerable: if the \(top-k\) ranking contains chunks below this threshold, the system assumes no answer exists. In the comparison, we evaluate similarity scores using unanswerable questions, which are then compared with scores from a \textit{control set} of answerable questions taken from \(Q_a\). The former indicates a loss of precision, as chunks are retrieved for questions with no answers. We evaluate RAG and GraphRAG on both question sets. \Framework\ is omitted because it scopes the KB using predefined semantics. This makes the search space equivalent to the empty set (\text{\O}), resulting in no loss of precision. In Figure \ref{fig:relevance}, a cumulative percentage is shown, indicating the relative amount of chunks retrieved at a given threshold. A high threshold will improve precision but comes at a proportional cost to recall, as relevant chunks with lower similarity are not retrieved.   

In Figure \ref{fig:relevance}, we observe that, at a threshold of 0.83, the RAG approaches retrieve the majority of chunks in the control set, while retrieving 19\%--28\% from the unanswerable set. Therefore, 0.83 is a sensible threshold that balances precision and recall. Nevertheless, retrieval precision remains low: on average, a quarter of the total corpus is retrieved as false positives. Interestingly, GraphRAG achieves higher precision because its semantics are used to scope the KB, resulting in fewer high-similarity hits. This suggests a trend in which more precise semantics yield higher precision, allowing for a lower threshold, which in turn improves recall. This is a striking result, as it underscores the main reason for \Framework's high overall performance. Traditional approaches cannot retrieve additional context beyond a given threshold because retrieval is constrained by the $top-k$ ranking. In contrast, by using semantics, \Framework\ can retrieve relevant chunks that fall far below this threshold, thereby improving recall without compromising on precision as semantically unrelated chunks are excluded. This suggests that \Framework\ mitigates the precision-recall dilemma inherent to traditional RAG approaches and IR \cite{recall_vs_precision}. Moreover, as non-existent entities return the empty set (\text{\O}), \Framework\ can indicate prior to retrieval whether a given \textit{fact} does not exist in the KB. This can mitigate unnecessary computations, which would otherwise introduce overhead, waste tokens, and lead to false responses and hallucinations \cite{hallucinations,ir_llms_survey,RAG_survey,rag_bloat}.


%% file: sections/06_discussion.tex
\input{sections/commands}
\section{Discussion}

\subsection{General Implications}

We observed that \Framework\ improves \textbf{(i) retrieval relevance} by considering semantically relevant chunks, and increasing recall by ranking these chunks more highly in the \(top-k\) ranking. Consequently, RAG systems can generate more accurate responses across different domains and KBs. Furthermore, we have shown that \Framework\ outperforms state-of-the-art approaches, such as GraphRAG ($+$15.8\%), even when only extending base RAG ($+$2.1\%). These gains enable more scalable systems without sacrificing accuracy or recall, as state-of-the-art approaches typically incur additional time and space complexity to achieve comparable or lower improvements.

The results indicate a substantial improvement in \textbf{(ii) retrieval efficiency} for \Framework, as embedding distances are calculated over smaller search spaces. This means that retrieval is faster in RAG systems and, consequently, that corpus size can scale substantially without slowing retrieval. The latter is explained by retrieval speed scaling with the average number of chunks per search space (e.g., per entity) rather than with the corpus size. This drastically improves scalability, facilitating larger systems that can answer a wider range of questions. This effectively gives LLMs more domain-specific knowledge, enabling them to provide better responses to users. Furthermore, as token usage is reduced through the smaller \(top-k\) ranking or avoided as unanswerable questions can be mitigated prior to retrieval, this can make the operation of RAG systems considerably more affordable. The smaller \(top-k\) ranking can also encourage the use of small open-source LLMs with reduced context windows, which is particularly useful for RAG systems that handle privacy-sensitive data, such as in the medical or financial domains. 

The evaluation suggests that \Framework\ avoids the loss of \textbf{(iii) retrieval precision} observed in traditional RAG approaches (19\%--28\%) when using unanswerable questions. A striking result is that the similarity threshold underlines a precision-recall dilemma, inherent to traditional RAG approaches. This dilemma is circumvented through semantic steering by identifying and excluding non-existent semantics prior to retrieval, facilitating lower threshold values, and improving recall. This may lead to new use cases for RAG systems, as designers are less constrained by the precision-recall dilemma. In addition, through semantic steering, it would be possible to use alternative retrieval, augmentation, or generation methods to further improve performance. This would, for example, allow the use of MMR instead of cosine similarity for open questions. 
 
Another benefit of \Framework\ is implicit in its design. Traditional RAG systems rely solely on user-generated prompts to steer IR. This means that steering in IR relies solely on variations in prompt formulation. Meanwhile, in \Framework, a user can additionally steer IR by specifying semantics. This directly improves user control by giving users two steering methods. Furthermore, by reducing retrieval turbulence caused by different user formulations, \Framework\ more closely aligns with user intent than traditional RAG approaches. This accommodates a wider range of user expertise levels. Nevertheless, there is a contrast: introducing semantic steering may require experience, since users need know-how to choose relevant semantics. Despite this gap, systems can guide users to select appropriate semantics through user interactions. As an example, a user may select a location on a map, which the system then uses as a semantic to apply \Framework. In addition, while we evaluated entity selection, alternative selection methods (i.e., see Figure \ref{fig:selection methods}) can further increase \Framework's flexibility, making it easier to adapt to diverse applications while still using the same underlying KB.

\vspace{-2mm}
\subsection{Generalizability}
In our evaluation, we focus exclusively on the comparison with traditional RAG and GraphRAG, despite the existence of many alternative approaches. To assess the generalizability of \Framework, we therefore map \textbf{15} distinct RAG approaches onto our framework and examine whether they can be fully expressed within its structure. This is achieved by explicitly defining, for each stage in \Framework, the method or mechanism used in the corresponding approach, as shown in Table \ref{tab:mapping}. Using this procedure, we obtained \textit{full coverage} across all evaluated RAG variants. Importantly, \Framework\ does not aim to replace existing RAG approaches, but rather to unify and extend them. This follows directly from its design: if the selection stage includes the full corpus ($U$), the search space \((S)\) reduces to the underlying RAG approach. As a result, \Framework\ can be applied to any RAG approach that adheres to the canonical retrieval, augmentation, and generation scheme, making it broadly applicable.

\vspace{-2mm}
\subsection{Limitations and Future Work}
\Framework\ relies on semantics to enable semantic steering, making its effectiveness dependent on the presence and quality of these semantics in the KB. In \Dataset, semantics are introduced through LLM-based chunk annotations, enabling entity-level specifications, at the cost of computational overhead and potential noise \cite{all_roads_lead_to_rome}; fine-grained entity semantics may not always be available or practical, as they require chunk-level annotations. Nevertheless, \Framework\ can also operate on higher-level semantic information, such as document sources, dates, or categories, which are far less resource-intensive to obtain as they rely on document-level annotation while still substantially reducing the search space. In future work, it is important to investigate alternative semantic selection methods, which remain underexplored in our evaluation and present further opportunities (see Figure \ref{fig:selection methods}). In particular, the use of \textit{manual selections} may shed light on the relation between user control and experience.

The evaluation was conducted on \Dataset, as no other benchmark currently exists for semantic steering in RAG. Consequently, the reported results may not directly generalize to other knowledge bases or application domains. In addition, \Dataset\ uses LLM-generated semantics to construct the selection space. As \Framework\ relies on the quality of these semantics, inaccuracies in the automated annotations may reduce its effectiveness. Nevertheless, LLM-generated semantics are widely used in practice, making \Dataset\ representative of many existing knowledge bases used for RAG \cite{all_roads_lead_to_rome}. Future work should evaluate \Framework\ across additional benchmarks as they become available, as well as investigate how different annotation strategies influence the performance of semantic steering.

In the evaluation, we used \textit{cosine similarity}, a widely used retrieval method in RAG approaches. However, other retrieval methods exist that, for example, employ re-ranking to improve chunk relevance. The use of semantic steering is applicable across a wide range of retrieval methods, since the selection space forms a new corpus over which retrieval is applied as is. Nevertheless, the performance of \Framework\ may vary across different retrieval methods, and the improvements shown in the evaluation will vary across different pipelines. Evaluating \Framework\ in combination with alternative retrieval methods may provide further insight into how semantic steering interacts with these approaches. In particular, it may prove to be a more resource-efficient alternative or complement to more computationally expensive retrieval techniques, while providing comparable or higher improvements in retrieval performance.

The current evaluation of \Framework\ focuses primarily on the retrieval stage. However, \Framework\ includes several other stages, such as the augmentation and generation stage, that could be explored in future work. These stages may highlight additional benefits, as semantic steering can be used to employ alternative augmentation or generation strategies that best fit user questions. For example, GraphRAG augments its prompts with relations, which are then used to answer relational questions during the generation stage \cite{graphrag}. Therefore, semantic steering could lead to holistic pipeline changes by combining traditional IR paradigms with RAG. 

While this work evaluates \Framework\ at the level of RAG approaches, some existing RAG systems implicitly implement forms of semantic steering through user interactions that influence the canonical RAG stages \cite{tian2026ragexplorervisualanalyticscomparative}. These mechanisms are often not explicitly modeled as semantic steering, but effectively induce similar behaviour at the system level. \Framework\ can serve as a unifying abstraction for these interactions, enabling a more systematic description, comparison, and design of such systems. Future work should therefore explore how \Framework\ can be applied at the system level to unify and analyse different RAG systems.


%% file: sections/07_conclusion.tex
\input{sections/commands}
\section{Conclusion}
In this paper, we investigated the impact of introducing semantic steering and adding selections to traditional RAG approaches through \Framework. Traditional approaches rely on overhead in space and time complexity to improve performance in the \colorunder{retrieval}{Retrieval}, \colorunder{augmentation}{Augmentation}, and \colorunder{generation}{Generation} stages. Furthermore, said approaches perform poorly on precision and recall, as chunks with high similarity scores are naively added to the context window, even when they lack semantic relevance to the underlying question, thereby bloating retrieval. \Framework\ mitigates these effects drastically by introducing a \colorunder{selection}{Selection} stage that makes retrieval and generation far more efficient, precise, and accurate. In our evaluation, we determined that computational overhead across the entire canonical RAG process can be reduced by leveraging semantics, avoiding automated entity tagging, and reducing the search space by orders of magnitude. Moreover, compared to state-of-the-art RAG approaches, a noticeable improvement in retrieval relevance is observed due to the removal of semantically irrelevant chunks from the search space. In addition, \Framework\ provides a solution to the precision-recall dilemma by simultaneously avoiding loss of precision while improving recall. Overall, these benefits allow for more scalable systems. Besides performance metrics, \Framework\ offers a way to increase user control, allowing users to steer the inference process using semantics while reducing turbulence. This makes \Framework\ a powerful tool for aligning user intent with machine interpretation. 